\documentclass[12pt,preprint]{aastex}

\def\phs{ph~cm$^{-2}$~s$^{-1}$} 
\def\phsk{ph~cm$^{-2}$~s$^{-1}$~keV$^{-1}$} \shortauthors{XX et al.}
\shorttitle
{The Crab Nebula with \textit{INTEGRAL} SPI}

\begin{document}

\title{The High Energy Emission of the Crab Nebula from 20 keV to 6 MeV
with \textit{INTEGRAL} SPI \footnote{
Based on observations with INTEGRAL, an ESA project with instruments and science data centre
funded by ESA member states (especially the PI countries: Denmark, France, Germany, Italy, 
Spain, and Switzerland), Czech Republic and Poland with participation of Russia and USA.} }

\author{E. Jourdain and J. P. Roques}
\affil{CESR, Universit\'e  de Toulouse [UPS]  and CNRS, UMR 5187, 9 Av. du Colonel Roche, BP 44346, 31028 Toulouse Cedex~4, France}

\author{\it Received  ; accepted  }


\begin{abstract}
The SPI spectrometer aboard the INTEGRAL mission   observes regularly the
Crab Nebula since 2003. 
We report on  observations distributed over  
5.5 years and investigate the variability of the intensity and spectral shape
of this remarkable source in the hard X-rays domain up to a few MeV.  While 
single power law models give a good description
in the X-ray domain (mean photon index   $\sim$ 2.05) and MeV domain (photon index   
$\sim$ 2.23), crucial information are contained in the  evolution of the slope
with energy between these two values. This study has been carried out trough
individual observations and long duration ($\sim$ 400 ks) averaged spectra. The stability of
the emission is remarkable and excludes a single power law model. The slopes 
measured  below and above 100 keV agree perfectly with the last values reported in 
the X-ray and MeV regions respectively, but without  indication of  a localized
break point. This suggests a gradual softening  in the  emission around 100 keV 
and thus  a continuous evolution rather than an actual change in the mechanism parameters.
In the MeV region, no significant deviation from the proposed power law model is
visible up to 5-6 MeV. 
   
Finally, we take advantage of the spectroscopic capability of the instrument to seek for
previously reported spectral features in the covered energy range with  negative results
for any significant cyclotron or annihilation emission on 400 ks timescales. \\
Beyond the scientific results, the performance and reliability of the SPI
instrument is explicitly demonstrated, with some details about the most appropriate
analysis method. 
\end{abstract}      


\maketitle

\section{Introduction}
The Crab is a major source for the high energy domain, where it
represents one of the brightest and uniquely stable source. At least two spectral components
 have to be considered in the observed emission,
 one from the pulsar itself and one from 
its surrounding nebula. It has been shown (Kuiper et al., 2001) that at least up to 1 MeV, 
the pulsed emission is  harder than
the nebula (non pulsed)  one  and that its importance
grows with energy, from a few \%
around 1 keV to 20-30 \% around 1 MeV.  While both  spectra can be 
roughly
 described by a power law, its slope changes with 
energy (broken power law).
In this highly magnetized environment, the main photon production mechanism is  
 synchrotron radiation by charged particles, with  Compton collisions which scatter
 primary photons in the very high  energy domain (E $\sim$ 1 GeV). The emitted spectrum reflects
thus
simultaneously the distribution(s) of the charged particles and the geometry of the magnetic field.


A review of the main historical observations in  X- and $\gamma$-rays   can be found 
in Ling and Wheaton (2003). These authors present  the results obtained by the \textit{CGRO} BATSE
instrument in the 30 keV-1.7 MeV energy band. The 1-2  weeks averaged spectra can be described by 
a variable broken power law with average photon indices of $\sim $ 2.1 and $\sim$  2.4 respectively 
below and above an energy break close to 100 keV. Moreover, a hardening above
670 keV is proposed to be potentially related to an additional component observed 
in the first \textit{CGRO} COMPTEL channels up to 3 MeV. 
  
The analysis presented here is based on \textit{INTEGRAL} SPI  data in the 20 keV - 8 MeV
energy band. The main 
characteristics of the SPI telescope are the spectroscopic capability
of the germanium detectors and the good sensitivity achieved
over more than 2 decades in energy
with a unique instrument. Since the Crab has not been used  as a standard candle
to adjust SPI energy transfert  matrix,  SPI  is an ideal tool
to perform a robust study of the emission of this source. 
Our goal is to investigate the overall shape of the hard X-ray emission
as well as the presence of potential spectral features linked to  annihilation 
and cyclotron processes,
 since these points remain open in spite of numerous observations.
 
\section{Instrument and Observations}

The \ INTEGRAL (INTErnational Gamma-Ray Astrophysics Laboratory) observatory 
is an ESA mission 
launched in 2002, October 17.
One of its 2 main instruments is the spectrometer, SPI (Vedrenne et al., 2003), which 
consists of an array of 19 high purity Germanium (Ge) detectors, with a  geometrical surface area of
508 cm$^{2}$ and  a thickness of 7 cm. The detector plane temperature is maintained at $\sim$ 80 K
to reach an energy resolution between 2 and 8 keV from 20 keV to 8 MeV. The association of a
coded mask together with the germanium crystals leads to an angular resolution of $2.6^\circ$ (FWHM)
over a $30^\circ$ field of view. A 5-cm thick BGO shield  (ACS, Anti-Coincidence Shield)
protects the telescope from charged particles and photons originating
 outside the field of view.
The first instrument in-flight performance is given in Roques et al. (2003). \\

Usually, each 3-day orbit is filled with roughly one hundred of exposures lasting 30-40 minutes.
During an observation dedicated to a given source, the pointing direction varies around the target by steps of $\sim 2^\circ$ within 
a $5 \times 5$ square or a 7-point hexagonal pattern. 
 This dithering procedure (see Jensen et al., 2003, for details)
 is used to optimize the
imaging capabilities of  both IBIS (the second main INTEGRAL instrument) 
 and SPI. This approach is crucial for SPI since it increases the number of measurements (equal, 
for one exposure, to the number of detectors, see equation (1))
on a same sky region, eliminating  ghost images and  permitting
a robust background and source  fluxes estimation.

\subsection{Instrument Management}
 
 The SPI telescope has been carefully calibrated before its launch (Atti\'e et al., 2003), ranging from
 individual calibration up to full instrument calibrations.
Each step has been reproduced and compared to Monte-Carlo
simulations. This provides confidence in the simulation tools and 
in our understanding of the instrument.
The simulation tool has then been used to construct the response matrix.
 This matrix has been split in two parts, one 
reflecting the geometry (ray tracing), the other taking into account the
energy redistribution (Sturner et al., 2003). The information has been
 put in IRF and RMF  files
respectively, for 51 energy values and 95 angles in both directions.
 Specific interpolation subroutines allow to recover the information for
 any value of energy and incident direction.\\
 Due to 2 detector failures, the 1st one (detector N$^\circ$2) during 
revolution 140 (December 6th, 2003),
the second one (detector N$^\circ$17) after 
revolution 214 (July 17th, 2004), SPI data have to be analysed  
with  background fields and IRF matrices according to the detector plane 
configuration (spi\_irf\_grp\_0021.fits, spi\_irf\_grp\_0022.fits, spi\_irf\_grp\_0023.fits,
for 19, 18 or 17 detectors, respectively, the RMF information being unaffected).
 Moreover, data will be added only within 
 periods corresponding to the same configuration.\\
 Note that a 3rd failure (detector N$^\circ$5) 
occured on 2009, 19 February, after the last revolution considered in  the present analysis.

The excellent energy resolution of the germanium  crystals makes it crucial to
determine very accurately the channel-energy relation for each detector and each revolution 
to  ensure reliable results for spectral studies.\\
For the low energy chain (0-2 MeV), we use 6 background lines between 23.438 keV and 1763.367 MeV to build 
 an energy-channel relation, E=f(c), through a  function of the form: \\
  $  E(c)=A0 \times Ln(c)+A1+A2 \times c+A3 \times c^2+A4 \times c^3 $ \\
  where the five free parameters (Ai)$_{i=0,4}$, are adjusted for each detector and each revolution.\\
For the high energy chain (2-8 MeV), a simple linear relation based on 2 lines (2754.028 and 7415.60 keV)
was used (2 free parameters).\\
Second order drifts
due to dependence on the detectors temperature on a few hours timescale have not been included.
This can result in small discontinuities in the final spectra at the edge of strong
background lines, but doesn't influence the final scientific conclusions.

A second issue has been investigated in order to use data above 1 MeV.
There, the energy output of the electronic chain is contaminated by  high energy 
particles, which saturate the electronic and generate false triggers around 
1.4 MeV (see SPI team document by Wunderer, 2005). 
Fortunately, a second electronic chain ("PSD") operates in parallele from
the fast 
preamplifier output and generates an independent trigger for photon energies
 between 650 keV and 2.2 MeV. 
The "PSD"
electronic box (originally designed to perform Pulse Shape Discrimination)
has an efficiency of 85\% on its global energy domain due to a higher dead time but is
not affected by the saturation problem.
We thus use the trigger signal  issued from the PSD electronic 
to confirm the
reality of the events between 650 keV and 2.2 MeV and complete a clean analysis
in the whole energy range covered by our instrument (20 keV-8 MeV). 
We have consequently  built background patterns identically formed with "PSD" tagged 
events only, in the corresponding energy range.
The appropriate factor (0.85) is
applied to the data and error bars to take into account the
additionnal dead time related to PSD event selection.

\subsection{Data Set}


 Crab is used as a reference target to monitor
 or calibrate instruments. During the INTEGRAL mission, regular campaigns 
to measure Crab are planned, (roughly twice a year).
  Since these observations are carried out mainly for
performance monitoring or calibration purposes, various exotic pointing schemes have been realized.
To ensure results as robust as possible, we restricted our set of data to 
the standard dithering patterns, optimized for the SPI telescope observations.
We have analysed 13 observations dedicated to this source  from 2003, February 19th to 2008 September 28th, 
corresponding to 10 periods generally spaced by $\sim$ 6 monthes (see Table~1).
After data analysis and cleaning, we obtained
1.2 $\times$10$^6$ seconds of effective observing time.
 
The stability of the instrument (within 5\%) has been checked over the mission timescale 
(Jourdain \& Roques, 2009), making us confident to sum data to obtain better
signal to noise ratio.
We have conducted our analysis by first analysing individual revolutions, and then 
building 3 long duration averaged spectra for 3 periods  (see Table 1) : \\
Sum 1 corresponds to the beginning of the mission (19 detectors
 alive) with 3 revolutions   and 402 ks of useful duration. \\
Sum 2 groups 7 observations (of 30 to 70 ks each) spaced by 6 monthes 
 from 2004 September to 2007 September for a total useful duration of 325 ks (17 detectors
 alive).\\
Sum 3 corresponds to a specific long campaign dedicated to the Crab,
performed in March and September 2008, with 3 entire revolutions for 462 ks (17 detectors
 alive). 


\subsection{Analysis Method}
 
The flux extraction in the count space is based on a model fitting algorithm using $\chi^2$ minimization, with 
a sky model consisting of $N_{s}$ sources at their theoritical positions. 
For a given energy bin, E, the SPI data, integrated during a time interval (or pointing) p,
 can be expressed with the general formula:
  
\begin{equation}
  D_{p} = \sum_{i=1}^{N_s} {M_{p,i} \times S_{p,i} } + B_{p} 
 \label{equation} 
\end{equation} 
\\
Where $D$, $M$, $B$ are $N_{d}$  elements vectors (one element per detector, 
$N_{d}$ = 19, 18 or 17), and represent respectively:
\\
$D_{p}(E)$: data (number of counts) measured on the detector plane for the pointing p.\\
$S _{p,i}(E)$:	 intensity of source i, during the pointing p.\\
$M_{p,i}(E)$ : 	SPI spatial/geometrical response (IRF) for the source i for the direction of pointing  p.\\
$B_{p}(E)$  : 	background counts measured on the detector plane for the pointing p.\\

For the Crab observations, the sky model contains only this source, since the contribution of 
any other source can be considered as negligible (i.e. A0535+262 doesnot exceed 15 mCrab during the
 considered periods).  
Nevertheless, this system is solvable only if some external (a priori) information  is introduced.\\
The main information we can introduce concerns the background component. From empty field observations,
the  relative factors  between detectors, U[d], are measured to describe the background maps  on the
detector plane . For each energy band, $B_{p}$ in equation~(\ref{equation}) is thus reduced to $\beta$(p)*U[d], 
with only one free parameter per pointing, the global normalisation factor $\beta$(p) of a fixed vector, 
d being the detector number.\\
We thus consider a set of $ N_{p}$ equations to treat simultaneously $N_{p}$ pointings.
The variability of the background  intensity, $\beta$(p), can  be constrained, since it is not expected to vary on the
pointing ( $\sim$ 2 ks)  timescale. By looking at the count rates measured in the anti-coincidence
 system, it can be seen that the background intensity is in general stable on the orbit (3 days)  
timescale. 
The  spectra have thus been built with the background 
normalisation $\beta$(p) allowed to vary on the revolution timescale ($\sim$ 2.5 days) while the source 
is considered as constant. Consequently, for an observational period encompassing $N_{p}$  pointings and $N_{r}$  revolutions, 
the number of free parameters is  $N_{r}$ (background intensities) + 1 (the Crab intensity), 
while the number of data is $N_{p}$*$N_{d}$ (with $N_{p} \sim 10-100  N_{r}$).

At the end of the flux extraction procedure, $\chi^2$ for individual energy channels 
are checked to detect any  problems. Except at low energy where the high level of
statistic surpasses our instrument knowledge precision (with  source detection levels greater than
100 $\sigma$) , all $\chi^2$ values are acceptable.

At this stage, a count spectrum and its associated response matrix (i.e. for the Crab nebula position 
and a given pointing sequence) are stored.


\section{Spectral Analysis}

From the count spectra and the corresponding matrices, we have reconstructed the  incident photon spectra 
through a spectral fitting procedure (Xspec11 tools).
 
Some non-statistical features appear in the lower channels due to the uncertainties on the energy response, which drops off 
dramatically in this domain, and threshold effects. We have thus excluded the  first channels
( $E <$ 23.5 keV) from the fit process. No systematic has been introduced in order to keep data as free
as possible of any  subjective  information.

\subsection{Analysis of individual observations}

Considering the 13 observations presented in Table~1, the spectral variability
 of the Crab  emission has been investigated at two levels: 
\begin{itemize}
\item
The potential evolution of the spectral shape in terms of one or two slopes model,
 as observed by BATSE (Ling \& Wheaton, 2003) 
\item
The stability of the parameters of the powerlaw(s).
\end{itemize}

The fits have been performed on 79 logarithmic channels between 23.5 keV and 1 MeV,
with single power law and broken power law models. 
In all cases, the 2 slopes model
provides a significant improvement in the $\chi^2$ value when compared to a single power law.
However, a  degeneracy between the break energy and the slopes (the low energy
one essentially) prevents us from constraining clearly the parameters. Indeed, a 
gradual steepening of the spectral emission toward high energies is probably closer
to the reality than the broken power law model.
This pushed us to test the parameters  variability through a broken powerlaw
model with the energy break fixed to 100 keV. This specific energy break has been chosen because it
allows easy comparisons with other observational results and is very close to the values reported
 from BATSE data (Ling \& Wheaton, 2003).\\    
 The fitted photon index values are displayed in Table~1.
They are  all compatible with photon indices of 2.07   and 2.24   below and above 100 keV 
respectively, demonstrating  a remarkable stability of the spectral shape emission between 23 keV and 1 MeV,
even though this analytical law doesn't represent a physical mechanism.\\
An important point, for comparison with other data sets, is the spectrum normalisation.
 The 100 keV flux in the last column  of Table~1 is the normalisation given by the broken powerlaw model, with
energy break fixed to 100 keV.
To obtain a model independent flux, we have made a local (82-118 keV) fit with a power law model, 
and obtained systematically lower values but always strikingly stable. (see Column~5 of Table~1).

\subsection{Analysis of long duration averaged spectra}
The three summed spectra encompassing more than 300 ks allow  us to better constrain the spectral shape, 
over an energy domain extended up to 6 MeV (7 channels more).
The  broken power law model with the break energy fixed to 100 keV gives
the same indices than for the individual spectra, but we can try to learn more by letting free the
 break energy. 
A simultaneous fit of the three spectra  converges toward  
 photons indices of  2.04 and 2.18, respectively below and above a break energy of 62 keV.
Figure 1 displays the 3 averaged spectra together with this broken power law fit.
However, it is clear that these values are 
strongly drawn by the low energy high signal to noise ratio. When adding 
some systematics to the data, we observe  that the fitted energy break
increases with the  value of the systematic, with correlated changes in the fitted slopes.
For example, a reasonable value of 1\% of systematic ends up in a "banana" 90\%  confidence region  
from 70 to 90 keV for the energy  break,
and from 2.05 to 2.08 for the first slope, the second one remaining above 2.18.
In fact,  we can see it as a region of more or less equivalent solutions, since 
we stress again  that  such a 2 slope model is an analytical description
of a probably smooth evolution. This can be illustrated by using a formula 
where the slope is continuously varying with Log(E) as proposed by Massaro et al. (2000) 

$F(E)=A \times E^{a+bLog(E/E_0)} $\\
A simultaneous fit of the three spectra gives 

a=1.79 $\pm$ 0.02

b=0.134 $\pm$ 0.01

A= 3.87 \phsk\  \\
with $E_0$ fixed to 20 keV.\\
We then get  $\chi$ = 421 for 253 dof, clearly better than  $\chi$ = 453 for 252 dof
obtained for the broken powerlaw model. To provide complete information, fluxes and errors of the three 
spectra are given in Table 2.\\
Note that the continuous curvature prevents any safe extrapolation outside the
considered energy range while representing very well the SPI data.\\

\section{Discussion}


The hard X-ray domain is important for the study of the Crab Nebula emission since
it joins the X-ray band where a photon index of $\sim$ 2-2.1 is known from many measurements
to the MeV and above domain  where observations are much rarer but indicate an increase of the photon index (2.2-2.3). 
The shape of this spectral evolution contains information on the main parameters of the emission itself, magnetic field value and
charged particles distribution. The disagreement between the most recent results limits any firm conclusion:
While the spectrum measured by the PDS instrument (15-300 keV; Kuiper et al. 2001) aboard 
the BeppoSAX mission  is compatible with a single  slope power law model (photon index of 2.14)
found by the GRIS balloon experiment (20-700 keV; Bartlett et al. 1994), the detailled 
analysis performed by Ling \& Wheaton (2003) with the BATSE data (35 keV-1 MeV)  concludes that the $\sim $10 days
averaged spectra are better described with a slope varying from 2.1 to 2.4 around a break at
$\sim $ 100 keV. Moreover, these authors report   a hardening above $\sim$ 700  keV, which could be
an additionnal structure superimposed to the powerlaw component.

From the analyses based on a large set of SPI data, we confirm the first finding  of 
BATSE, namely the slope softening in the hard X-ray emission of the total spectrum.
Our results show  a quite satisfying agreement with the BATSE ones as the two slopes and energy break 
values can be considered as the same. Moreover, the 40 keV fluxes are quite similar 
(4.5 $\times 10^{-3}$ \phsk\ for BATSE
to compare to 4.4 $\pm 0.1\times 10^{-3} $  
\phsk\ obtained from SPI data).
However, we can not confirm  the variability observed
on a 10 days timescale nor the emission in  excess of the powerlaw observed  above 700 keV 
(see section~\ref{Comptel}). 

\subsection{Comparison with X-ray results}
The X-ray slope  (above 1.5 keV to avoid absorption effects) has a standard value of
2.1 but recent study with XMM data (Kirsch et al. 2006) reports a value of 2.046 for the
total (nebula + pulsar) spectrum. This "mean" slope summarizes in fact strong variations of
the photon index inside the nebula structure (from 2.0 to 2.4) and during the phase of the pulsed emission
(from 1.5 to 1.75), which are similarly contained in our
total spectrum. The agreement between the  slope  we have obtained in the low energy part of SPI spectra 
and this mean X-ray value confirms that the same mechanisms  underlie the emission
from soft to hard X-rays.

 Concerning the spatial variations of the emission (very well studied below 10 keV), a first
 result from a balloon-borne telescope in the 22-64 keV energy range (Pelling et al. 1987) points out 
  the interest of spatially resolved (below 30'') data 
in the hard X-ray domain  for future instruments. It is the only way to access  the topological 
information and to link  a particular  region with the macroscopic parameters
 deduced from observations  (magnetic field value, for example).

\subsection{Comparison with results at higher energy}\label{Comptel}
The main observations for comparing our results in the MeV region have been performed with COMPTEL. 
 Kuiper et al. (2001)
present a complete analysis of pulsed and non pulsed emission. 
They report a clear evolution of the pulse morphology explained by a three-component model,
with fraction of pulsed over non pulsed emission presenting a maximum ($\sim 30\%$) around 1~MeV.

The high energy slope deduced from SPI data (2.24 $\pm$ 0.04) is quite consistent with the values deduced from COMPTEL 
data (photon index of 2.35 for the pulsed emission, 2.23 for the nebula). However, while it 
implies  a  pronounced break around 1 MeV in the pulsed emission (see Kuiper et al, 2001),
it corresponds to a more continous curvature of the nebula spectral emission, pointing out
noticeable differences in the emitting processes.
 However, in this region, the total emission as measured by SPI does not exibit any particular feature. 
More precisely, the hardening observed
above 700 keV in the BATSE data and  possibly related to some similar structure in the first COMPTEL 
channels  cannot be confirmed. Figure 2 shows the high energy part of our 3 averaged spectra. 
No deviation from the high energy power law can be observed while the flux remains unchanged within errors.
 Sligthly positive residuals are present above 2-3 MeV, but
well above the BATSE excess and  their significance is too low to reject a
statistical nature.

\subsection{Search for annihilation and cyclotron features}

The excellent energy resolution of the SPI detector plane makes it the best instrument to 
seek for (narrow) spectral features in the observed emission. Concerning the Crab Nebula,
several detections have been reported of two different natures, attributed to cyclotron emission 
in one hand, at E $\sim$ 75 keV, and to annihilation  in the other hand, around 400 keV and above 
(see Owens (1991) for a summary). For the cyclotron emission, detections correspond to fluxes 
between 0.3 and 1 $\times 10^{-2}$ \phs\ while the most constraining 3 $\sigma$ upper limit is
 6.2 $\times 10^{-4}$ \phs. As regard the "annihilation" mechanism, photon production varies between
  1 $\times 10^{-4}$ \phs and 7.2 $\times 10^{-3}$ \phs. Variabilities (in energy, flux and 
  possibly phase region) point out the complexity of the physics contained in the Crab emission.

  
We have looked for the presence of similar structures  in the SPI data by
 re-building spectra with 5 keV width channels. 
 In each averaged spectrum, no significant excess has been found, in the mentionned energy domains.
 Considering the summed spectra, this leads to 4 $\sigma$ upper limits of 2 $\times 10^{-4}$ \phs\ in a  5 keV width band, around 75 keV  
while  they are between 1.5 and 2 $\times 10^{-4}$ \phs\ for a 3 $\sigma$ significance in the 400-600 keV domain.
Concerning a broader feature,  upper limit  values are between 2 and 3.3 $\times 10^{-4}$ \phs\ for a 25 keV band  
centered on 511 keV, for the accumulated spectra.

In conclusion, no feature corresponding to cyclotron or annihilation emission  on
timescales of 400 ks has been detected, for the total Crab emission. 
  
\section{ Conclusions}

We have used the \textit{INTEGRAL} SPI data to study the spectral emission of the Crab from 20 keV 
to a few MeV. As preliminary remarks, it is worth to mention that the SPI
telescope benefits from an "absolute" calibration, in the sense that its transfert matrices
are based on ground calibrations and Monte-Carlo simulations, and therefore independent from Crab observations,
which are often used  for calibrations of other X-and $\gamma$-ray instruments. Moreover, 
 details on some analysis aspects  which 
require   particularly careful handling are explicited .

Beyond the good agreement of our results with previous works, particularly with the analysis
of BATSE data by Ling \& Wheaton (2003), we point out the impressive stability of the
total Crab emission from 20 keV to a few MeV. The spectral parameters have been 
found very stable on the 6 years timescale  and  rule out
a single powerlaw model.  Broken power law 
 models give a good description of the SPI data even though the energy break 
cannot be firmly constrained, probably because the transition from the low energy 
slope to the high energy one is smooth.

Three mean spectra have been built in order to achieve a better statistic 
(total useful duration of  $\sim$400 ks) and investigate deeply the spectral shape, 
including the MeV region.
 Using a broken power law model, the best fit procedure
converges toward  slope values  of 2.07 $\pm$ 0.01 and 2.23 $\pm$ 0.05 for a break fixed at
100 keV. When letting free the energy break, the fit procedure converges toward a value of 62 keV,
while the power-law indices become 2.04 and 2.18 respectively. 
These values are close to those reported from instruments in
adjacent energy bands, pointing out a coherent view of the wide band emission.

Another important result is  the continuous curvature in the 100 keV region,
we can describe by the analytical law
\begin{equation}
  F(E)=3.87 \times E^{1.79 + 0.134 Log(E/20)}~ph~cm^{-2}~s^{-1}~keV^{-1}
 \label{fit} 
\end{equation}  
 valid in the 23 keV-6 MeV domain, with E expressed in keV .\\
This gradual steepening provides information about the acceleration
mechanism(s) and charged particle  distributions participating to the X/$\gamma$-ray production
and  gives opportunities to test models and learn more about emission parameters.
However, the pulsed/non-pulsed fraction  as well as the spatial distribution
have to be investigated in more details in order to better understand the global nebula emission in hard X-rays. 
All these ingredients plays a crucial role in the global/averaged shape observed in this domain
and have be considered in the model in order to explain the observed spectral emission.

 \section*{Acknowledgments}  The \textit{INTEGRAL} SPI project has been completed under the
  responsibility and leadership of CNES.
   We are grateful to ASI, CEA, CNES, DLR, ESA, INTA, NASA and OSTC for support.

\begin{figure}
\plotone{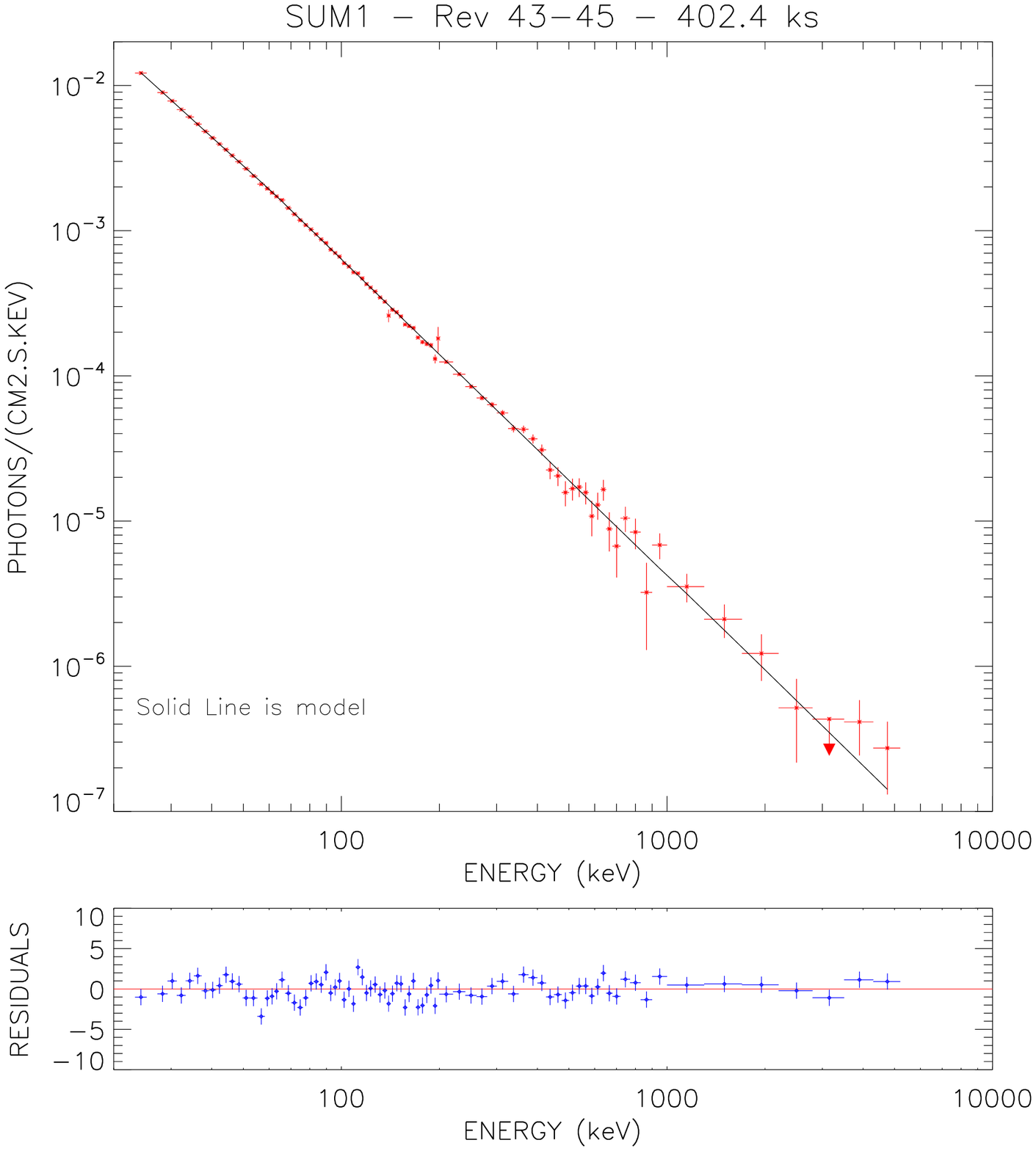}
\end{figure}

\begin{figure}
\plotone{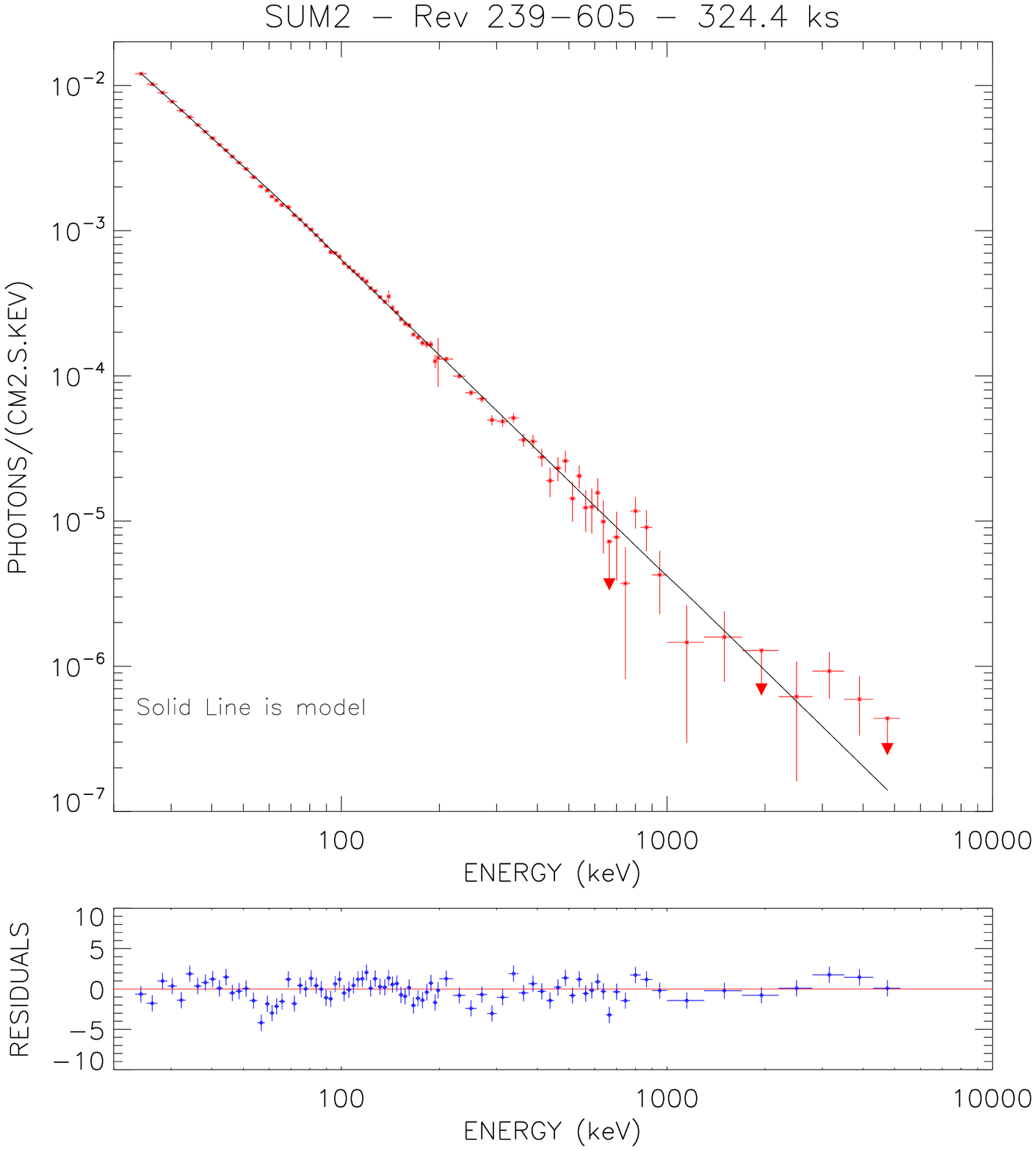}
\end{figure}

\begin{figure}
\plotone{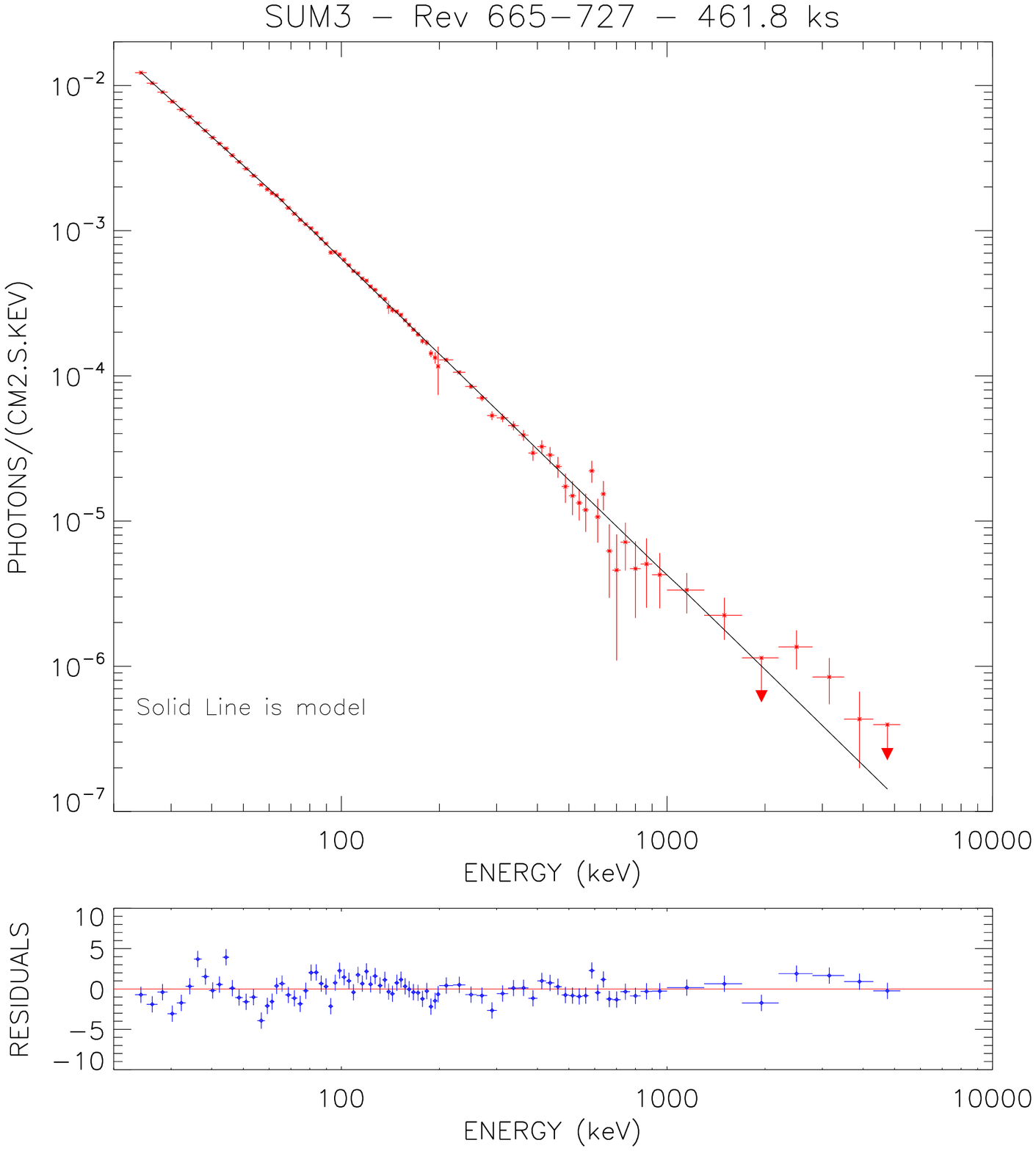} 
\caption{ Crab averaged spectra for the three epochs (sum1, sum2, sum3, see Tab.~1).
Solid line represents the best fit broken power law model obtained from a simultaneous fit.
No systematic has been added to the data.}
\end{figure}

\begin{figure}
\plotone{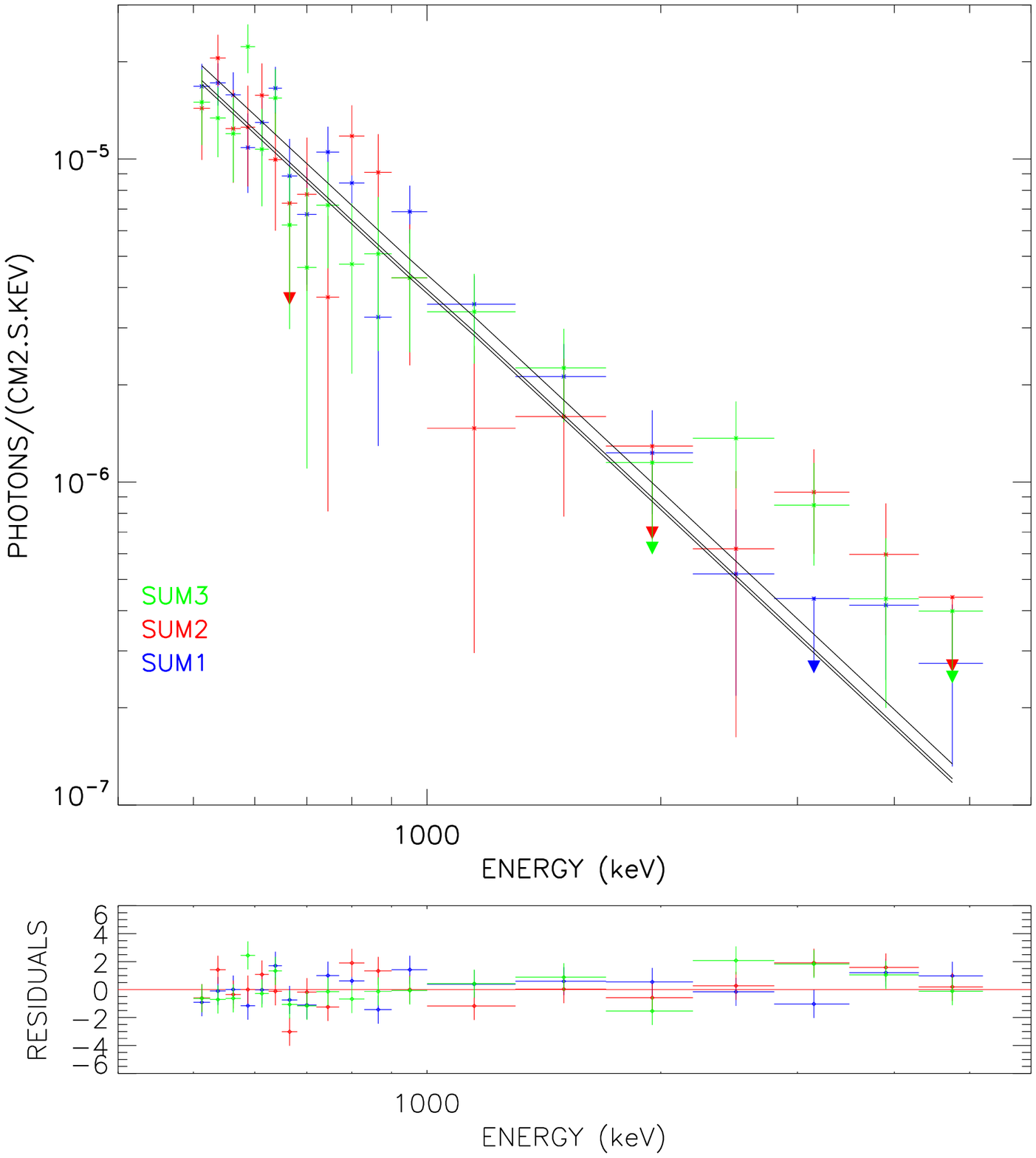}

\caption {Crab  spectra for three epochs (sum1, sum2, sum2) above 500 keV; Solid lines
represent the high energy power law (photon index of 2.24).}
 
\end{figure}


\begin{deluxetable}{lccccccccc}
\tablewidth{0pt}
\tablecaption{Log of the \textit{INTEGRAL} SPI observations dedicated to the Crab used in this paper (standard dithering modes).}

\label{tab:revol}
\tabletypesize{\scriptsize}
\tablehead{
 \colhead{}\\
 
\colhead{revol }
&\colhead{Start} 
&\colhead{End} 
&\colhead{useful}
 &\colhead{$F_{100 keV}$}
&\colhead{$index1 $} 
&\colhead{$index2 $}
&\colhead {$F_{100 keV}$} \\
\colhead{number}&&&\colhead{duration}
 &\colhead{$\times 10^{-4}~ph~/$}
 &&&\colhead{$\times 10^{-4}~ph~/$}\\
&&&\colhead{ks}
 &\colhead{$~cm^{2}~s~keV$}
&&&\colhead{$~cm^{2}~s~keV$}

&\colhead{} \\
 
 }
\startdata 
 

 43 & 2003-02-19 16:06:05 &2003-02-21 15:59:22 & 139.94   & 6.3 $\pm$ 0.2    & 2.08 $\pm$   0.01  & 2.25 $\pm$ 0.03 & 6.5 $\pm$ 0.2 \\
 44 & 2003-02-22 04:06:15 &2003-02-24 12:10:57 & 156.22   & 6.5 $\pm$ 0.2    & 2.07 $\pm$   0.01  & 2.24 $\pm$ 0.03 & 6.6 $\pm$ 0.2  \\
 45 & 2003-02-25 03:58:54 &2003-02-26 16:51:08 & 106.22   & 6.5 $\pm$ 0.2  & 2.05  $\pm$   0.01 & 2.24 $\pm$ 0.03  & 6.8 $\pm$ 0.2\\
 Sum 1    & rev 43 & rev 45   & 402.38 & 6.45 $\pm$   0.1 &  2.07 $\pm$   0.01  &   2.24  $\pm$ 0.02  & 6.6 $\pm$ 0.1\\
239 &  2004-09-29 11:12:34&2004-09-29 22:38:05 &  31.14   & 6.4 $\pm$   0.2  & 2.07 $\pm$   0.02  & 2.33 $\pm$ 0.07 & 6.6 $\pm$ 0.2\\
300 &  2005-03-29 12:00:27&2005-03-30 01:18:30 &  38.72   & 6.3 $\pm$   0.2  & 2.08 $\pm$   0.02  & 2.23 $\pm$ 0.07 & 6.5 $\pm$ 0.2\\
365 &  2005-10-11 08:33:49  &    2005-10-11 19:11:44    &  30.51   &  6.6  $\pm$   0.2     & 2.05 $\pm$   0.02  & 2.31 $\pm$   0.07 & 6.8 $\pm$ 0.2\\
422 &  2006-03-28 17:28:11&2006-03-29 07:27:19 &  38.70   & 6.5 $\pm$   0.2  & 2.07 $\pm$   0.0   & 2.23 $\pm$ 0.07& 6.6 $\pm$ 0.2 \\
483$^*$ & 2006-09-29 01:40:14 & 2006-09-29 14:27:34  & 32.41    & 6.5  $\pm$   0.3      & 2.1 $\pm$   0.03  & 2.26 $\pm$   0.08 & 6.5 $\pm$ 0.2\\
541 & 2007-03-19 13:25:14 &2007-03-20 15:34:16 &  71.34   & 6.3 $\pm$   0.2  & 2.07  $\pm$  0.02  & 2.27 $\pm$ 0.08 & 6.4 $\pm$ 0.2\\
605 & 2007-09-27 00:21:34 &2007-09-28 06:09:51 &  81.56   & 6.4 $\pm$   0.2  & 2.07  $\pm$  0.02  & 2.24 $\pm$ 0.05& 6.6 $\pm$ 0.2\\
Sum 2   & Rev 239  &  Rev 605 & 324.4    &  6.35 $\pm$   0.1    &     2.07  $\pm$   0.01  & 2.25 $\pm$ 0.03 & 6.55 $\pm$ 0.1\\
665 & 2008-03-24 11:57:5 & 2008-03-27 00:06:30 &  146.44  & 6.6 $\pm$   0.2  & 2.07  $\pm$  0.01  & 2.27 $\pm$ 0.03& 6.7 $\pm$ 0.2 \\
666 & 2008-03-27 11:40:28& 2008-03-29 23:51:34 &  154.35  & 6.5 $\pm$   0.2  & 2.06  $\pm$  0.01  & 2.26 $\pm$ 0.03& 6.7 $\pm$ 0.2 \\
727 &   2008-09-25 22:34:01& 2008-09-28 11:06:32 &  161.02   &  6.5  $\pm$   0.2     &    2.07  $\pm$   0.01& 2.22 $\pm$ 0.03 & 6.7 $\pm$ 0.2\\
Sum 3 &  Rev 665 & Rev 727 &  461.81   &   6.5 $\pm$ 0.1  & 2.06$\pm$   0.01 & 2.25  $\pm$ 0.04& 6.7 $\pm$ 0.1\\

\enddata

\tablecomments{Columns 6 to 8 give the best fit parameters for a broken power law model
with the break energy fixed to 100 keV. The data have been fitted between 23.5 and 1 MeV (6 MeV
for summed spectra) with no systematic. Column 5 is the model independent 100 keV flux (see section 3.1).\\
$*$ Rev 483 has been performed with a 5x5 grid with 1$^\circ$ degree step instead of 2$^\circ$. 
}
 
\end{deluxetable}

\begin{deluxetable}{lcccccccc}
\tablewidth{0pt}
\tablecaption{Fluxes and errors for the three Crab spectra (sum1, sum2 and sum3) 
in the energy channels used in this analysis. The deconvolution is based on the
 powerlaw with energy dependent slope model (equation (2)).
Fluxes and errors are given in units of $ ph~/~cm^{2}~s~keV$}

\label{tab:flux}
\tabletypesize{\scriptsize}
\tablehead{
 \colhead{}\\
 
\colhead{$E_{min}$ }
&\colhead{$E_{max}$} 
&\colhead{Flux (sum1)} 
&\colhead{error (sum1)}
 &\colhead{Flux (sum2)}
&\colhead{error (sum2)} 
&\colhead{Flux (sum3)}
&\colhead {error (sum3)} \\
 
}
\startdata 

  23.25 &   25.25 & 1.22E-02 & 7.17E-05 & 1.20E-02 & 1.06E-04 & 1.22E-02 & 9.40E-05 & \\
  25.25 &   27.25 & 1.02E-02 & 4.25E-05 & 1.02E-02 & 6.06E-05 & 1.04E-02 & 5.32E-05 & \\
  27.25 &   29.25 & 8.94E-03 & 2.90E-05 & 8.91E-03 & 4.06E-05 & 8.99E-03 & 3.54E-05 & \\
  29.25 &   31.25 & 7.81E-03 & 2.52E-05 & 7.72E-03 & 3.48E-05 & 7.74E-03 & 3.02E-05 & \\
  31.25 &   33.25 & 6.82E-03 & 2.24E-05 & 6.72E-03 & 3.06E-05 & 6.83E-03 & 2.66E-05 & \\
  33.25 &   35.25 & 6.07E-03 & 2.02E-05 & 6.04E-03 & 2.77E-05 & 6.09E-03 & 2.40E-05 & \\
  35.25 &   37.25 & 5.42E-03 & 1.87E-05 & 5.34E-03 & 2.55E-05 & 5.50E-03 & 2.21E-05 & \\
  37.25 &   39.25 & 4.82E-03 & 1.74E-05 & 4.80E-03 & 2.37E-05 & 4.89E-03 & 2.06E-05 & \\
  39.25 &   41.25 & 4.35E-03 & 1.63E-05 & 4.33E-03 & 2.23E-05 & 4.37E-03 & 1.95E-05 & \\
  41.25 &   43.25 & 3.95E-03 & 1.56E-05 & 3.90E-03 & 2.14E-05 & 3.97E-03 & 1.87E-05 & \\
  43.25 &   45.25 & 3.61E-03 & 1.50E-05 & 3.58E-03 & 2.08E-05 & 3.68E-03 & 1.83E-05 & \\
  45.25 &   47.25 & 3.29E-03 & 1.51E-05 & 3.23E-03 & 2.09E-05 & 3.30E-03 & 1.83E-05 & \\
  47.25 &   49.75 & 2.98E-03 & 1.29E-05 & 2.94E-03 & 1.83E-05 & 2.98E-03 & 1.62E-05 & \\
  49.75 &   52.25 & 2.67E-03 & 1.47E-05 & 2.66E-03 & 2.16E-05 & 2.67E-03 & 1.92E-05 & \\
  52.25 &   55.25 & 2.38E-03 & 2.93E-05 & 2.33E-03 & 4.12E-05 & 2.39E-03 & 3.59E-05 & \\
  55.25 &   58.25 & 2.09E-03 & 2.02E-05 & 2.02E-03 & 2.90E-05 & 2.07E-03 & 2.56E-05 & \\
  58.25 &   60.25 & 1.95E-03 & 2.54E-05 & 1.89E-03 & 3.61E-05 & 1.92E-03 & 3.17E-05 & \\
  60.25 &   62.25 & 1.82E-03 & 2.56E-05 & 1.72E-03 & 3.66E-05 & 1.81E-03 & 3.22E-05 & \\
  62.25 &   64.25 & 1.72E-03 & 2.91E-05 & 1.62E-03 & 4.14E-05 & 1.75E-03 & 3.63E-05 & \\
  64.25 &   67.25 & 1.62E-03 & 3.21E-05 & 1.50E-03 & 4.50E-05 & 1.62E-03 & 3.92E-05 & \\
  67.25 &   70.25 & 1.43E-03 & 1.47E-05 & 1.45E-03 & 2.16E-05 & 1.43E-03 & 1.93E-05 & \\
  70.25 &   73.25 & 1.30E-03 & 9.05E-06 & 1.27E-03 & 1.32E-05 & 1.30E-03 & 1.17E-05 & \\
  73.25 &   76.25 & 1.18E-03 & 8.81E-06 & 1.19E-03 & 1.26E-05 & 1.19E-03 & 1.11E-05 & \\
  76.25 &   79.25 & 1.09E-03 & 8.36E-06 & 1.09E-03 & 1.19E-05 & 1.10E-03 & 1.05E-05 & \\
  79.25 &   82.25 & 1.02E-03 & 7.61E-06 & 1.02E-03 & 1.09E-05 & 1.04E-03 & 9.66E-06 & \\
  82.25 &   85.25 & 9.43E-04 & 7.71E-06 & 9.31E-04 & 1.11E-05 & 9.62E-04 & 9.81E-06 & \\
  85.25 &   88.25 & 8.71E-04 & 7.89E-06 & 8.58E-04 & 1.15E-05 & 8.79E-04 & 1.02E-05 & \\
  88.25 &   91.25 & 8.23E-04 & 8.42E-06 & 7.84E-04 & 1.24E-05 & 8.13E-04 & 1.11E-05 & \\
  91.25 &   94.25 & 7.41E-04 & 1.79E-05 & 7.12E-04 & 2.49E-05 & 7.07E-04 & 2.18E-05 & \\
  94.25 &   97.25 & 7.02E-04 & 1.09E-05 & 7.02E-04 & 1.58E-05 & 7.14E-04 & 1.40E-05 & \\
  97.25 &  100.25 & 6.64E-04 & 9.57E-06 & 6.64E-04 & 1.38E-05 & 6.86E-04 & 1.22E-05 & \\
 100.25 &  103.75 & 5.97E-04 & 9.64E-06 & 5.97E-04 & 1.37E-05 & 6.31E-04 & 1.21E-05 & \\
 103.75 &  107.25 & 5.67E-04 & 6.39E-06 & 5.60E-04 & 9.40E-06 & 5.78E-04 & 8.38E-06 & \\
 107.25 &  110.75 & 5.17E-04 & 6.10E-06 & 5.27E-04 & 8.99E-06 & 5.28E-04 & 7.99E-06 & \\
 110.75 &  114.25 & 5.09E-04 & 6.07E-06 & 4.99E-04 & 8.97E-06 & 5.10E-04 & 7.97E-06 & \\
 114.25 &  117.75 & 4.70E-04 & 6.06E-06 & 4.68E-04 & 9.01E-06 & 4.69E-04 & 8.03E-06 & \\
 117.75 &  121.25 & 4.29E-04 & 6.59E-06 & 4.48E-04 & 9.82E-06 & 4.54E-04 & 8.73E-06 & \\
 121.25 &  124.75 & 4.06E-04 & 6.26E-06 & 4.03E-04 & 9.67E-06 & 4.14E-04 & 8.65E-06 & \\
 124.75 &  129.25 & 3.82E-04 & 5.47E-06 & 3.86E-04 & 8.21E-06 & 3.93E-04 & 7.34E-06 & \\
 129.25 &  133.75 & 3.47E-04 & 5.62E-06 & 3.50E-04 & 8.54E-06 & 3.57E-04 & 7.67E-06 & \\
 133.75 &  138.25 & 3.25E-04 & 6.87E-06 & 3.26E-04 & 1.08E-05 & 3.39E-04 & 9.78E-06 & \\
 138.25 &  141.25 & 2.61E-04 & 2.58E-05 & 3.53E-04 & 3.57E-05 & 2.99E-04 & 3.11E-05 & \\
 141.25 &  145.75 & 2.86E-04 & 7.73E-06 & 2.95E-04 & 1.34E-05 & 2.85E-04 & 1.22E-05 & \\
 145.75 &  150.25 & 2.76E-04 & 5.38E-06 & 2.75E-04 & 8.22E-06 & 2.79E-04 & 7.38E-06 & \\
 150.25 &  154.75 & 2.58E-04 & 5.16E-06 & 2.47E-04 & 7.77E-06 & 2.64E-04 & 6.95E-06 & \\
 154.75 &  159.25 & 2.27E-04 & 5.39E-06 & 2.29E-04 & 8.09E-06 & 2.43E-04 & 7.23E-06 & \\
 159.25 &  163.75 & 2.21E-04 & 5.77E-06 & 2.24E-04 & 8.57E-06 & 2.26E-04 & 7.63E-06 & \\
 163.75 &  169.25 & 2.15E-04 & 4.75E-06 & 1.94E-04 & 7.19E-06 & 2.09E-04 & 6.44E-06 & \\
 169.25 &  174.75 & 1.85E-04 & 5.09E-06 & 1.85E-04 & 7.99E-06 & 1.94E-04 & 7.19E-06 & \\
 174.75 &  180.25 & 1.72E-04 & 5.47E-06 & 1.70E-04 & 8.53E-06 & 1.75E-04 & 7.70E-06 & \\
 180.25 &  185.75 & 1.67E-04 & 6.10E-06 & 1.66E-04 & 8.92E-06 & 1.71E-04 & 7.95E-06 & \\
 185.75 &  191.25 & 1.63E-04 & 6.19E-06 & 1.66E-04 & 9.17E-06 & 1.44E-04 & 8.22E-06 & \\
 191.25 &  196.75 & 1.32E-04 & 9.13E-06 & 1.27E-04 & 1.35E-05 & 1.35E-04 & 1.20E-05 & \\
 196.75 &  199.75 & 1.82E-04 & 3.62E-05 & 1.34E-04 & 4.93E-05 & 1.17E-04 & 4.28E-05 & \\
 199.75 &  220.25 & 1.26E-04 & 2.81E-06 & 1.32E-04 & 4.25E-06 & 1.30E-04 & 3.83E-06 & \\
 220.25 &  240.25 & 1.04E-04 & 2.48E-06 & 1.01E-04 & 3.70E-06 & 1.07E-04 & 3.30E-06 & \\
 240.25 &  260.25 & 8.51E-05 & 2.47E-06 & 7.74E-05 & 3.71E-06 & 8.54E-05 & 3.32E-06 & \\
 260.25 &  280.25 & 7.11E-05 & 2.65E-06 & 7.03E-05 & 3.98E-06 & 7.13E-05 & 3.56E-06 & \\
 280.25 &  300.25 & 6.40E-05 & 2.69E-06 & 5.02E-05 & 4.04E-06 & 5.39E-05 & 3.61E-06 & \\
 300.25 &  325.25 & 5.61E-05 & 2.65E-06 & 4.91E-05 & 3.89E-06 & 5.20E-05 & 3.46E-06 & \\
 325.25 &  350.25 & 4.38E-05 & 2.48E-06 & 5.19E-05 & 3.67E-06 & 4.60E-05 & 3.27E-06 & \\
 350.25 &  375.25 & 4.33E-05 & 2.53E-06 & 3.67E-05 & 3.73E-06 & 3.95E-05 & 3.32E-06 & \\
 375.25 &  400.25 & 3.73E-05 & 2.63E-06 & 3.58E-05 & 3.87E-06 & 2.98E-05 & 3.45E-06 & \\
 400.25 &  425.25 & 3.13E-05 & 2.69E-06 & 2.79E-05 & 3.91E-06 & 3.30E-05 & 3.48E-06 & \\
 425.25 &  450.25 & 2.27E-05 & 3.06E-06 & 1.93E-05 & 4.40E-06 & 2.89E-05 & 3.89E-06 & \\
 450.25 &  475.25 & 2.07E-05 & 3.06E-06 & 2.35E-05 & 4.41E-06 & 2.41E-05 & 3.91E-06 & \\
 475.25 &  500.25 & 1.59E-05 & 3.11E-06 & 2.64E-05 & 4.50E-06 & 1.75E-05 & 3.99E-06 & \\
 500.25 &  525.25 & 1.69E-05 & 2.89E-06 & 1.45E-05 & 4.44E-06 & 1.52E-05 & 3.98E-06 & \\
 525.25 &  550.25 & 1.74E-05 & 2.54E-06 & 2.08E-05 & 3.71E-06 & 1.35E-05 & 3.30E-06 & \\
 550.25 &  575.25 & 1.59E-05 & 2.74E-06 & 1.26E-05 & 3.99E-06 & 1.21E-05 & 3.56E-06 & \\
 575.25 &  600.25 & 1.10E-05 & 3.00E-06 & 1.27E-05 & 4.35E-06 & 2.25E-05 & 3.87E-06 & \\
 600.25 &  625.25 & 1.31E-05 & 2.76E-06 & 1.59E-05 & 4.05E-06 & 1.09E-05 & 3.62E-06 & \\
 625.25 &  650.25 & 1.67E-05 & 2.73E-06 & 1.01E-05 & 4.00E-06 & 1.56E-05 & 3.56E-06 & \\
 650.25 &  680.25 & 8.97E-06 & 2.69E-06 & -1.5E-06 & 3.69E-06 & 6.33E-06 & 3.32E-06 & \\
 680.25 &  720.25 & 6.82E-06 & 2.68E-06 & 7.88E-06 & 3.91E-06 & 4.68E-06 & 3.56E-06 & \\
 720.25 &  770.25 & 1.06E-05 & 2.10E-06 & 3.79E-06 & 2.96E-06 & 7.30E-06 & 2.64E-06 & \\
 770.25 &  830.25 & 8.54E-06 & 2.03E-06 & 1.20E-05 & 2.92E-06 & 4.80E-06 & 2.60E-06 & \\
 830.25 &  900.25 & 3.29E-06 & 1.97E-06 & 9.23E-06 & 2.89E-06 & 5.17E-06 & 2.58E-06 & \\
 900.25 & 1000.25 & 6.97E-06 & 1.41E-06 & 4.36E-06 & 2.02E-06 & 4.36E-06 & 1.80E-06 & \\
1000.25 & 1300.25 & 3.61E-06 & 8.02E-07 & 1.50E-06 & 1.19E-06 & 3.43E-06 & 1.06E-06 & \\
1300.25 & 1700.25 & 2.16E-06 & 5.59E-07 & 1.63E-06 & 8.28E-07 & 2.30E-06 & 7.40E-07 & \\
1700.25 & 2200.25 & 1.26E-06 & 4.44E-07 & 5.08E-07 & 6.61E-07 & 1.38E-08 & 5.88E-07 & \\
2200.25 & 2800.25 & 5.32E-07 & 3.08E-07 & 6.37E-07 & 4.71E-07 & 1.40E-06 & 4.21E-07 & \\
2800.25 & 3500.25 & 1.17E-07 & 2.23E-07 & 9.57E-07 & 3.41E-07 & 8.72E-07 & 3.05E-07 & \\
3500.25 & 4300.25 & 4.27E-07 & 1.76E-07 & 6.15E-07 & 2.69E-07 & 4.48E-07 & 2.42E-07 & \\
4300.25 & 5200.25 & 2.82E-07 & 1.47E-07 & 1.65E-07 & 2.27E-07 & 1.01E-07 & 2.05E-07 & \\

\enddata
\end{deluxetable}

\end{document}